# Analysing Scientific Collaborations of New Zealand Institutions using Scopus Bibliometric Data

Samin Aref
Department of Computer Science and
Te Pūnaha Matatini
University of Auckland, New Zealand
sare618@aucklanduni.ac.nz

David Friggens
Ministry of Business Innovation
& Employment, Wellington
New Zealand
david.friggens@mbie.govt.nz

Shaun Hendy
Department of Physics and
Te Pūnaha Matatini
University of Auckland, New Zealand
shaun.hendy@auckland.ac.nz

## ABSTRACT

Scientific collaborations are among the main enablers of development in small national science systems. Although analysing scientific collaborations is a well-established subject in scientometrics, evaluations of scientific collaborations within a country remain speculative with studies based on a limited number of fields or using data too inadequate to be representative of collaborations at a national level. This study represents a unique view on the collaborative aspect of scientific activities in New Zealand. We perform a quantitative study based on all Scopus publications in all subjects for more than 1500 New Zealand institutions over a period of 6 years to generate an extensive mapping of scientific collaboration at a national level. The comparative results reveal the level of collaboration between New Zealand institutions and business enterprises, government institutions, higher education providers, and private not for profit organisations in 2010-2015. Constructing a collaboration network of institutions, we observe a power-law distribution indicating that a small number of New Zealand institutions account for a large proportion of national collaborations. Network centrality concepts are deployed to identify the most central institutions of the country in terms of collaboration. We also provide comparative results on 15 universities and Crown research institutes based on 27 subject classifications.

## CCS CONCEPTS

• **Information systems** → **Digital libraries and archives**; • **Applied computing** → *Document management and text processing*; • **General and reference** → Surveys and overviews;

## KEYWORDS

Big data modelling, Scientific collaboration, Scientometrics, Network analysis, Scopus, New Zealand

This is an author copy of the paper. The publisher's verified version of the paper can be accessed on https://doi.org/10.1145/3167918.3167920

ACM acknowledges that this contribution was authored or co-authored by an employee, contractor or affiliate of a national government. As such, the Government retains a nonexclusive, royalty-free right to publish or reproduce this article, or to allow others to do so, for Government purposes only.

## 1 INTRODUCTION

There is a growing body of literature that recognises the importance of scientific collaboration in economic development [1]. The scientific collaborations can be analysed based on bibliometric data using network analysis tools and techniques [2]. The main objective of analysing scientific collaboration is to gain an understanding of how knowledge flows between authors [3, 4], institutions [5] and countries [6, 7]. It also helps quantifying research performance measures with a focus on the collaborative aspect of research [8].

Scientific collaboration is seen not only as a performance measure, but also a representation an entity outreach and connections to other entities. Some studies focus on collaborations within a country to compare researchers or institutions and facilitate national research policy development. Perc analysed collaboration at the level of individuals in Slovenia [9] and similar study has been undertaken for Turkey [10]. Collaborations can be investigated between different countries. Park et al. investigated collaborations between China and South Korea using bibliometric data [11]. Nguyen et al. analysed collaborations of Vietnam with several other countries [12].

The university-industry collaboration has been investigated extensively as an essential connection between institutions of a science system. Abramo et al. investigated the university-industry collaboration in Italy [13] and found that university researchers collaborating with industry have a higher research performance. Investigating collaborations between specific types of institutions in a country usually requires an analysis of research outputs that represent a collaboration tie between the two types of institutions [13]. Yoon and Park investigated the collaboration between South Korea universities, industry, and government using network analysis tools and techniques on patent data [14]. The intermediate step of using network analysis to study scientific collaboration is evaluating joint outputs of authors affiliated with different types of institutions. Each bibliometric record of such nature represents a visible research connection that can be aggregated for evaluating collaboration at a national level.

It is important to point out that three types of collaborations have long existed in economic development literature and are relevant to the role of scientific collaboration; the triadic relationship between academia, industry and, the government is referred to as The Triple Helix. The term was coined by Etzkowitz and Leydesdorff [15] to refer to the shift from a dyadic industry-government relationship in an industrial society to a complex hybridisation of elements from academia, industry and government in a knowledge society. This shift is a result of innovation dynamics that support economic development.

The main contribution of this study is quantifying different types of scientific collaboration in New Zealand (NZ). This requires studying co-publications of all pairs of New Zealand institutions to evaluate the current engagement level between them. Research collaboration among various institutions is critical to policy development as it facilitates evaluating the current state of collaboration and helps identifying capacities for improvement in different fields of research. While a few global studies exist that provide some general observations on New Zealand scientific collaboration, a specific study on New Zealand academia, government and corporations' collaboration has never been undertaken. Following the triple helix concept, we investigate collaboration among all New Zealand institutions that have a publication in a scientific database within a specified time range.

## 2  OVERVIEW OF NZ COLLABORATIONS

Scopus and Web of Science (WoS) are two competing bibliometric databases of academic publications. Scopus is owned by Elsevier and accompanied by SciVal[1] an analytics service for Scopus data. WoS is maintained by Clarivate Analytics (formerly Thomson Reuters).

In what follows, basic results from Scopus and WoS that are related to New Zealand in comparison to Small Advanced Economies Initiative countries[2] are discussed:

1. University-Industry Research Connections (UIRC) series of studies [16] based on the WoS data
2. An implementation of a snowball metric [17] in SciVal based on Scopus data

### 2.1  UIRC Report 2014

The most recent UIRC report [16] uses University-Industry Co-publication (UIC) as an indicator of scientific performance in various countries. Using 2009-2012 WoS data, the report shows that three New Zealand universities (University of

Auckland, University of Canterbury, and Massey University) have an average level of overall UIC while the other two investigated (Victoria University of Wellington and University of Otago) have a medium-low level of overall UIC. These evaluations are based on a scale of high, medium-high, average, medium-low, and low UIC in comparison with other universities listed in 2014 edition of Leiden University Ranking [3] . The overall scores of the five New Zealand universities across seven disciplines can be found in UIRC 2014 report [16]. The report also demonstrates that New Zealand has an average overall UIC score which is above Small Advanced Economies Initiative countries like Singapore and Israel, but below Denmark, Switzerland, Finland and Ireland. Note that, UIRC 2014 does not include Crown Research Institutes (CRIs), which are government owned research laboratories accounting for a considerable amount of New Zealand scientific collaborations. Methodological details of UIRC 2014 can be found in a study of public–private collaboration [18].

### 2.2  SciVal Academic-Corporate metric

SciVal is a bibliometric analysis service based on the Scopus data providing research performance of 7500 research institutions worldwide (at the time of access). SciVal implements the Snowball Metrics [17]; The Academic-Corporate Collaboration Snowball metric is calculated based on Scopus data in 2011-2016. Numerical values for 15 New Zealand universities and CRIs range from 0.4% to 1.8% suggesting that they are comparably active in collaboration with corporations. Note that SciVal's institutional mapping and classification is not complete. It has better coverage of the larger organisations and misses many small ones. Notably it only has one commercial NZ institution (Fonterra). The overall Academic-Corporate Collaboration metric for New Zealand is the lowest among Small Advanced Economies Initiative countries.

## 3  MATERIALS AND METHODS

In order to quantify New Zealand research collaborations at a national level, we consider all New Zealand institutions that have a publication in Scopus within the six-year time window of 2010-2015. We standardise and classify thousands records of collaborations based on 2010-2015 publications to a list of institution pairs and their reciprocal number of joint publications[4] [19].

We used a full extract of Scopus [20] limiting to NZ publications between 2010 and 2015 which covers a considerable portion of the scientific publication relevant for our purposes. This was combined with Ministry of Business Innovation & Employment (MBIE) internally developed mapping of Scopus Affiliation IDs to NZ institutions. Scopus uses an automated process to cluster unstructured affiliation text and apply its internal Affiliation IDs, but this is conservative and results in multiple IDs per institution. MBIE's manually compiled mapping groups together all Affiliation IDs for each NZ institution (for example there are 81 Affiliation IDs for University of Auckland) and assigns a category (see below). Some affiliations are missing or are not identifiable as specific institutions (eg independent researchers), and in some cases there are data errors incorrectly identifying an affiliation as being from NZ; these were excluded. Data analysis was performed using the R language (version 3.4.2) and RStudio software (version 1.1.383). Gephi version 0.9.1 is used for network visualisations.

While UIRC 2014 contains 5 NZ universities and there are only 37 NZ institutions listed in SciVal (at the time of access), our study comprises over 1500 New Zealand institutions. This study mainly focuses on collaboration measures of 15 New Zealand universities and Crown Research Institutes (CRIs) listed below that are expected to account for a large proportion of New Zealand publications in Scopus:

(1) AgResearch (New Zealand Pastoral Agriculture Research Institute), (2) AUT University (Auckland University of Technology), (3) ESR (Environmental Sciences Research), (4) GNS Science (the Institute of Geological and Nuclear Sciences), (5) Landcare Research, (6) Lincoln University, (7) Massey University, (8) NIWA (National Institute of Water and Atmospheric Research), (9) Plant and Food Research (New Zealand Institute for Plant and Food Research), (10) Scion (New Zealand Forest Research Institute Limited), (11) University of Auckland, (12) University of Canterbury, (13) University of Otago, (14) University of Waikato, and (15) Victoria University of Wellington.

We use the numbers in the list above to refer to the 15 institutions later in the paper. The key measure to be used is the collaboration record which is the number of joint publications for pairs of institutions in Scopus within the specified time range. An author with affiliations to two or more institutions is not counted as a collaboration.

We adopt the triple helix concept [15] and use a classification system with four institution classes. The classification is a mapping between Scopus affiliation IDs, standard institution names, and one of the four categories: (1) Business enterprise, (2) Private not for profit (PNP), (3) Government, and (4) Higher education.

Business enterprises are institutions registered in New Zealand Companies Office Register [21].

Private not for profit institutions include institutions classified as building society, charitable trust, contributory mortgage broker, credit union, friendly society, incorporated society, industrial & provident society, limited partnerships, other bodies, overseas issuer, participatory security, retirement villages, superannuation scheme, or unit trust in New Zealand Companies Office other Registers [22].

Government class comprises of Crown Research Institutes, central government institutions, local government institutions, other government institutions, schools, public hospitals, and district health boards.

Finally, the higher education class includes private training establishments, universities, polytechnics, institutes of technology, independent research organisations, and education providers classified as wānanga.

For pairs of collaborating institutions, the collaboration records measure equals total number of joint publications. However, when collaboration records are aggregated for a specific institution, this equality does not hold. Consider a publication that has AUT, ESR, and GNS as the affiliations of its three distinct authors. Such a publication will be counted three times: once as a collaboration record of AUT-ESR, once for ESR-GNS, and once for AUT-GNS. Recalling that CRIs and universities belong to government and higher education categories respectively, when we aggregate the collaborations of AUT with the government sector, the publication contributes two to the collaboration records between AUT and the government sector.

## 4 RESULTS AND DISCUSSION

### 4.1 New Zealand Collaboration Network

In this subsection, we discuss overall network properties of a scientific collaboration network in which nodes are institutions and edges represent collaborations between them. Four institution categories define node types and the collaboration records values are used as weights on the edges.

The network has 1511 nodes and 4273 edges (network density is 0.004). Degrees of the nodes follow a power-law distribution with many institutions having less than 30 collaborators while in the tail distribution a few institutions have hundreds of collaborators.

Fig. 1 shows a visualization of the network in which size of the nodes are proportional to their weighted degrees. Nodes are coloured respective to their classifications. Business enterprises are shown in red, government institutions in green, higher education institutions in blue, and PNP organisations in purple.

The network in Fig. 1 is made of one relatively large component as well as 21 two-node components and 2 three-node components representing 23 groups of collaborating institutions isolated from the rest of network.

Like other collaboration networks [3,4], NZ collaboration network has the small world property. It has an average clustering coefficient of 0.53. For the giant component, the average path length (degree of separation) is 2.75 which much shorter than the network diameter which is 6.

66.8% of the institutions in Fig.1 are business enterprises. Proportions for PNPs, government institutions, and higher education institutions are 18.5%, 11.1%, and 3.6% respectively. The average unweighted degree of the network is 5.66 which represents the average number of collaborators for a given institution. The average number of collaboration records per institution (average weighted degree) is 38.13.

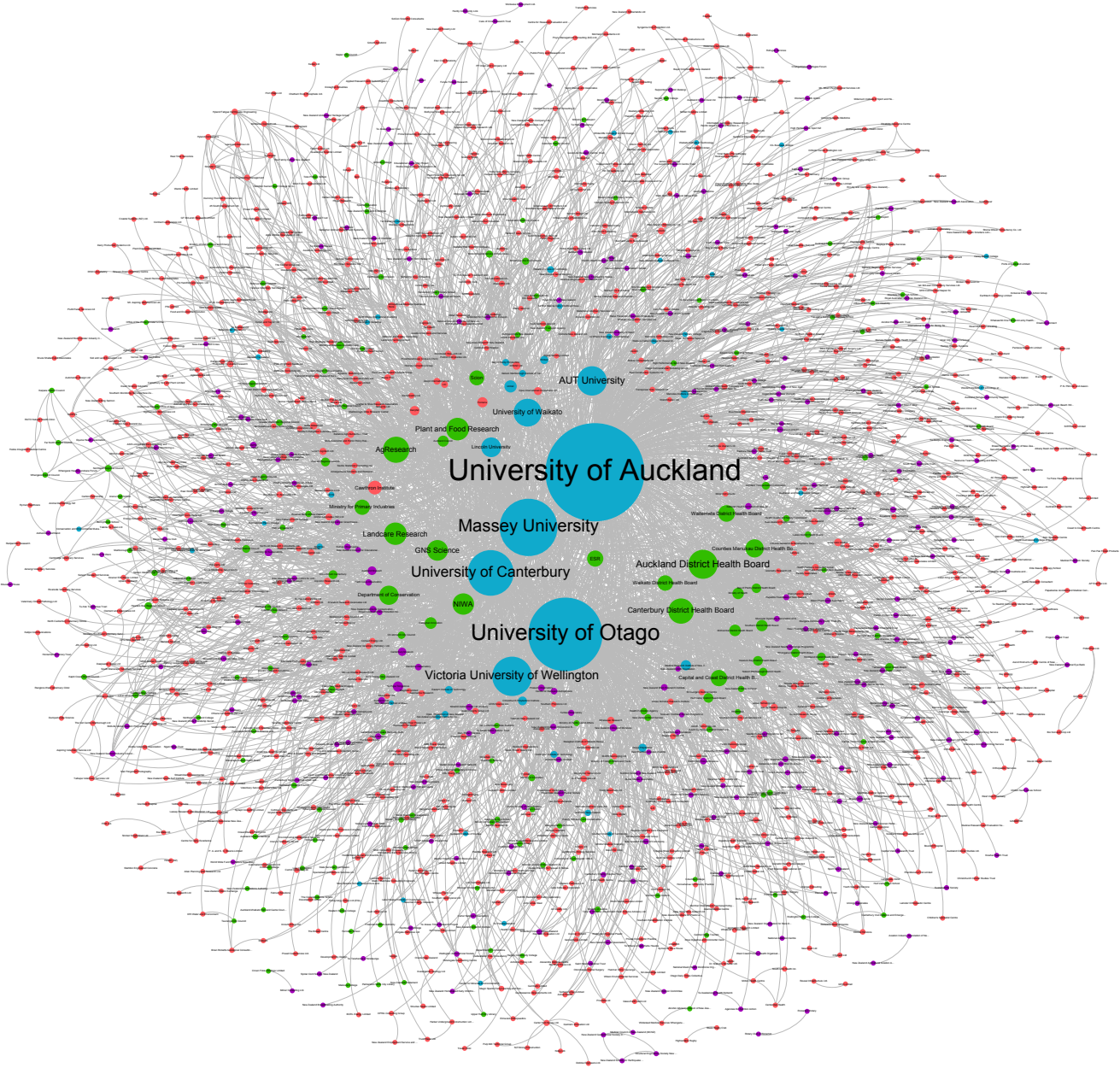

**Figure 1:** New Zealand Scientific collaboration network visualised as a weighted network (image credit © Samin Aref)

## 4.2 Induced Subgraphs of Crown Research Institutes

In what follows, smaller parts of the network, commonly referred to as ego networks, are illustrated. Ego networks are induced subgraphs of neighbours centered at a specific node.

Figs. 2-8 represent the induced subgraphs of 7 CRIs and their collaborators. Each induced subgraph represents the main institution located in the centre with a distinctive node size. The other nodes represent the collaborating institutions that are all connected to the central node and coloured respective to the classification. Thickness of the edges is proportional to the number of joint publications between the collaborating institutions. Business enterprises are shown in red, government institutions in green, higher education institutions in blue, and PNP organisations in purple.

From the induced subgraph in Fig. 2 we observe that AgResearch has strong collaboration ties to higher education institutions. There are many business enterprises collaborating with AgResearch as shown in Fig. 2. The ESR induced subgraph shows strong collaboration ties to higher education institutions. There are only a few collaborating PNPs in the induced subgraph of ESR in Fig. 3, while government category seems to account for many ESR collaborators. Like subgraphs of AgResearch and ESR, GNS Science induced subgraph and that of Landcare Research show strong collaboration ties to higher education institutions. Fig. 4 shows that a few PNPs collaborate with GNS Science, while many PNPs have collaboration with Landcare Research as illustrated in Fig. 5.

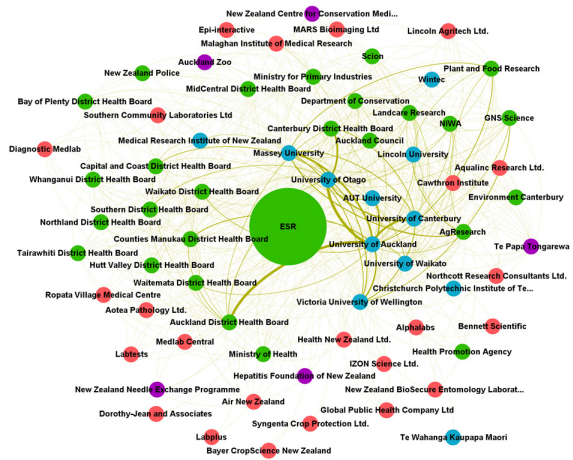

**Figure 3: ESR induced subgraph**

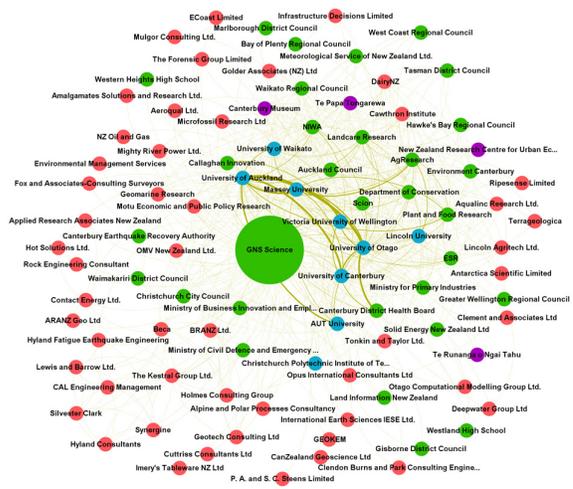

**Figure 4: GNS Science induced subgraph**

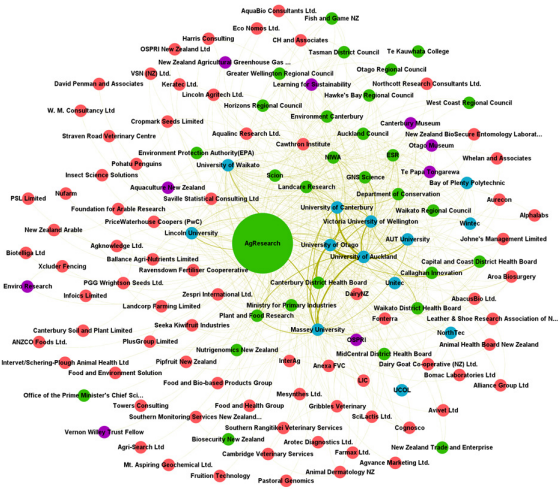

**Figure 2: AgResearch induced subgraph**

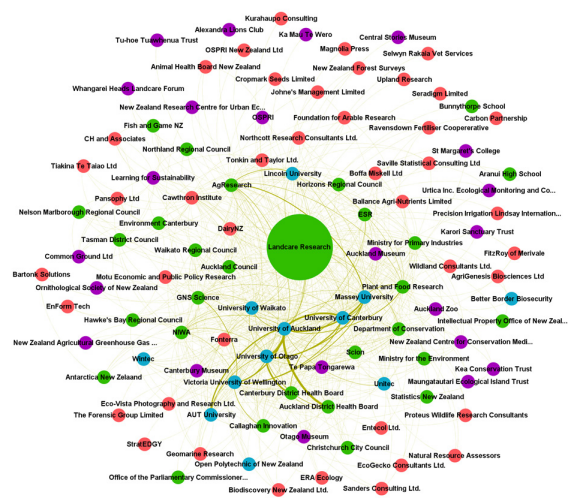

**Figure 5: Landcare Research induced subgraph**

Fig. 6 indicates that NIWA collaborates with all four types of institutions and most actively with higher education institutions. There are many PNPs collaborating with NIWA compared to other CRIs. Fig. 7 demonstrates relatively many business enterprises collaborating with Plant and Food Research. Strong ties to higher education institutions are visible for Plant and Food Research induced subgraph. The number of Scion collaborators is less than most other CRIs as evident in Fig. 8 which only shows one PNP collaborator.

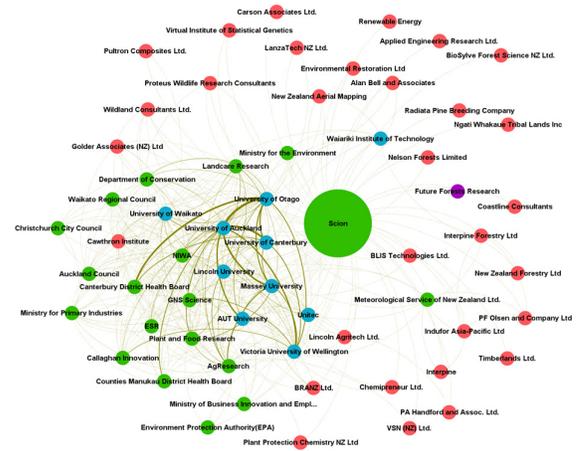

**Figure 8: Scion induced subgraph**

### 4.3 Centrality Analysis of the Network

The rest of this Section provides quantitative results on two centrality analyses that determine the most central institutions of the network.

Considering New Zealand collaboration network as a map representing the connections between institutions, it would be insightful to find the most important nodes of the network [22,23]. Betweenness [23] and eigenvector centrality [24] are standard network analysis tools that provide a quantitative measure of centrality for nodes of a given network. Betweenness centrality captures the importance of a node in a network based on its role of connecting other nodes. It measures how often a specific node appears on a path between two other nodes [23]. Eigenvector centrality contains an aggregate of a node's degree and its neighbours' degrees summed up based on a decreasing weight of distance to the neighbour [24].

Figs. 9-10 show ten most central institutions based on betweenness and eigenvector centrality measures respectively.

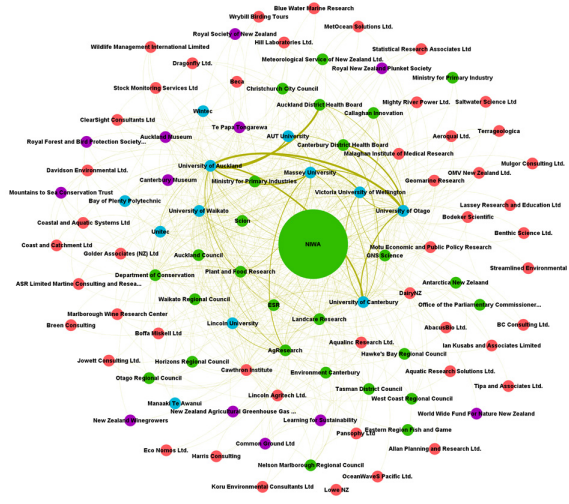

**Figure 6: NIWA induced subgraph**

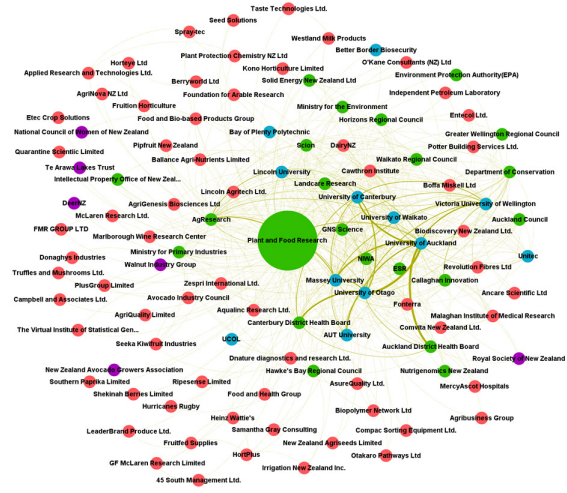

**Figure 7: Plant and Food Research induced subgraph**

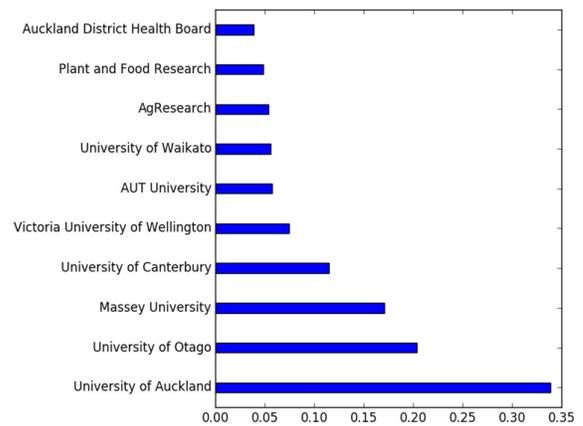

**Figure 9: Central institutions based on betweenness centrality**

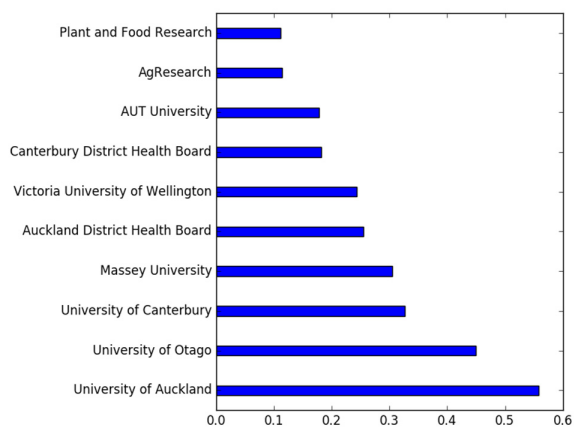

**Figure 10:** Central institutions based on eigenvector centrality

The most striking observation to emerge from Figs. 9-10 is that not all NZ universities are among the 10 most central institutions in terms of collaboration. Instead some district health boards and CRIs can be seen in Figs. 9-10. Recall that in Fig. 1, the eight universities were not exactly the eight largest nodes of the network. This means that based on degree centrality, some district health boards and CRIs are more central than some NZ universities.

## 5    COLLABORATION RATIOS

In this Section, we analyse the ratios of collaboration for each of the 15 universities and CRIs. This section can be considered as an analysis of weighted degrees of the nodes.

The collaboration ratios are provided as proportions in Fig. 11 and total counts in Fig. 12. Numbers on the vertical axis refer to the 15 universities and CRIs as listed in Section 3. Purple, blue, green, and red colours in Figs. 11-12 represent collaboration with PNPs, higher education providers, government institutions, and business enterprises respectively.

Three faceted plots based on All Science Journal Classification[5] (ASJC) are provided in the appendix. They can be used for comparing scientific collaborations in different fields of research as well as comparing universities and CRIs based on their collaboration records in each ASJC subject.

A comparison of relative collaborations with the four institution categories can be made based on Fig. 11. Lincoln university (6) has the highest proportion of collaboration with government. Regarding collaboration with business enterprises, Scion (10) has the highest proportion. University of Canterbury (12) has the highest proportion of collaboration with PNPs.

Fig. 12 shows that University of Auckland has the highest collaboration count followed by University of Otago, Massey University, University of Canterbury, and Victoria University of

Wellington. This order corresponds to the five largest nodes in Fig. 1 (recall that, the collaboration record counts equal weighted degrees of the nodes).

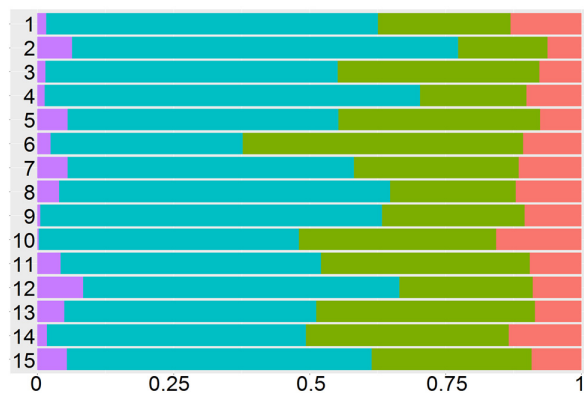

**Figure 11:** Collaboration records proportions for 15 universities and Crown research institutes

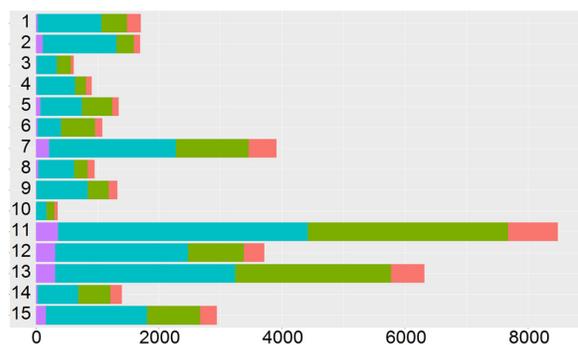

**Figure 12:** Collaboration records counts for 15 universities and Crown research institutes

## 6    CONCLUSIONS

This study investigated research collaborations among New Zealand institutions based on all Scopus publications from 2010 to 2015. Research collaborations were quantified by numerical measures based on joint publications of New Zealand institutions.

We have considered four classifications for the institutions, namely higher education, government, business enterprise, and private not for profit. The raw data containing thousands of Scopus affiliation IDs was categorised into standard institution names and classes defining nodes of a scientific collaboration network where collaborations are represented by weighted edges. The network is unique in its representative capability for research collaborations at a national level.

---

[5] A journal subject classification including 1 general and 26 specific subject classes: Agricultural and Biological Sciences, Arts and Humanities, Biochemistry Genetics and Molecular Biology, Business Management and Accounting, Chemical Engineering, Chemistry, Computer Science, Decision Sciences, Dentistry, Earth and Planetary Sciences, Economics Econometrics and Finance, Energy, Engineering, Environmental Science, Health Professions, Immunology and Microbiology, Materials Science, Mathematics, Medicine, Neuroscience, Nursing, Pharmacology Toxicology and Pharmaceutics, Physics and Astronomy, Psychology, Social Sciences, and Veterinary.

The centrality analysis, demonstrated in Figs. 9-10, indicated the most central institutions in terms of scientific collaboration. The quantitative results on collaboration records, illustrated in Figs. 11-12, shed light on collaborations between New Zealand universities/CRIs and different types of institutions at a national level. We have also used ASJC subjects to analyse research collaborations in 27 different fields of research whose comparative results can be found in the Appendix. It would be insightful as a future research to investigate whether the same observations, including the centrality analysis, hold true if other sources of bibliometric data are used.

While we focused on 15 universities and CRIs in New Zealand, the analysis was performed for over 1500 New Zealand institutions comprising of business enterprises, charitable trusts, union trusts, incorporated societies, and limited partnerships registered in New Zealand companies office register as well as central and local government institutions, schools, district health boards, private training establishments, polytechnics, institutes of technology, and independent research organisations.

This research has opened many avenues to be explored by more in-depth analysis on New Zealand bibliometric data. For one research direction, the analysis can be extended allowing for measures of research quality to play a role in evaluating collaborations. Field-normalised citation based measures [25] might be suitable candidates to be used as measures of research quality. Evaluating the potential improvement capacities across different disciplines would be another direction that can be taken from a policy development perspective. Observing a few institutions accounting for a large proportion of collaboration and most of the results confirming intuitive expectations, a third recommendation for future work is incorporating a measure of institution size to get the relevant measures per capita and use them for a better comparison of research collaboration performance of the institutions.

## ACKNOWLEDGMENTS

Andrew Marriott and Sam Holmes at Ministry of Business Innovation & Employment (MBIE) performed much of the work classifying New Zealand institutions. The first author would like to thank Peter Ellis and Franz Smith for their support. This work was partially supported by the following sources:
- Te Pūnaha Matatini 2016-2017 postgraduate internship program at Ministry of Business, Innovation & Employment
- The Computing Research and Education Association of Australasia partial support grant (CORE Student Travel Award)
- Callaghan Innovation grant number 9152 3705825

## A  APPENDIX

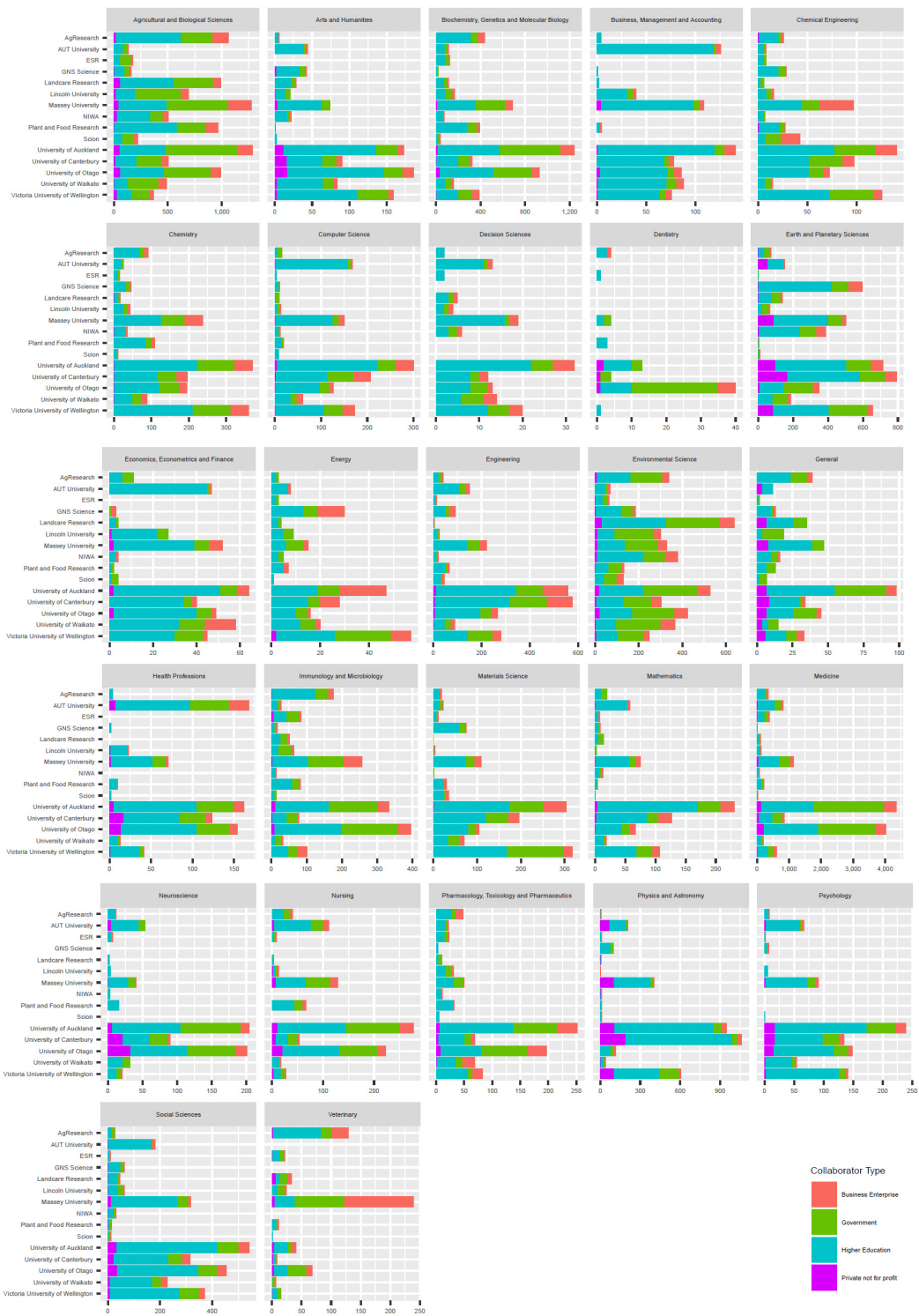

**Collaboration records count within New Zealand for each ASJC fields**

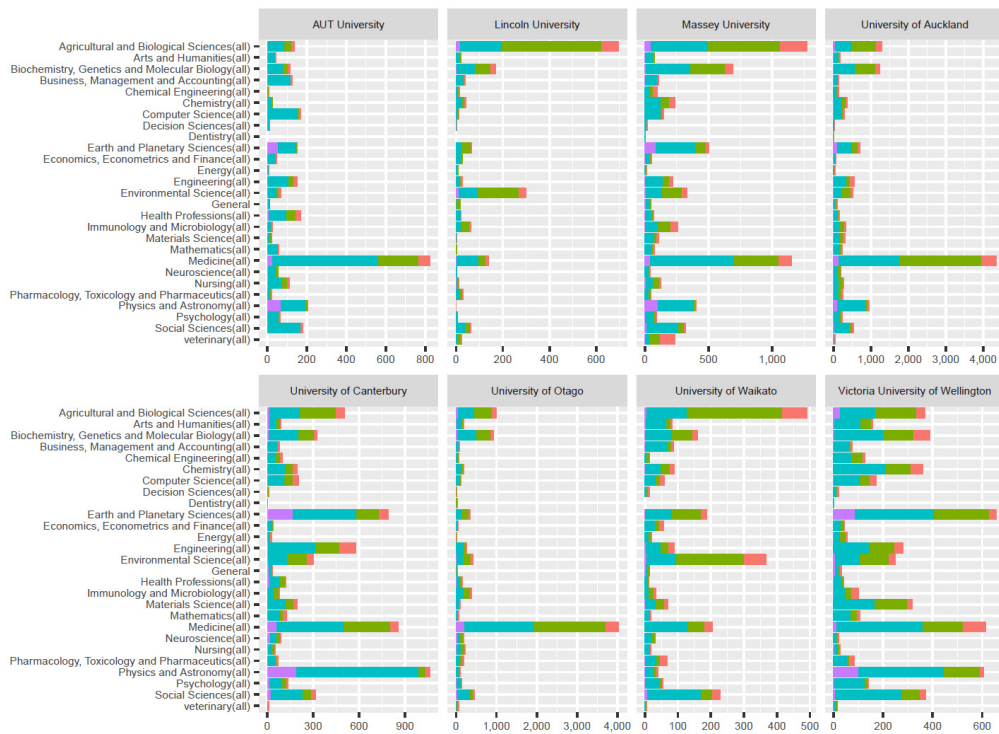

**Comparison of collaboration records of 8 New Zealand universities based on ASJC field**

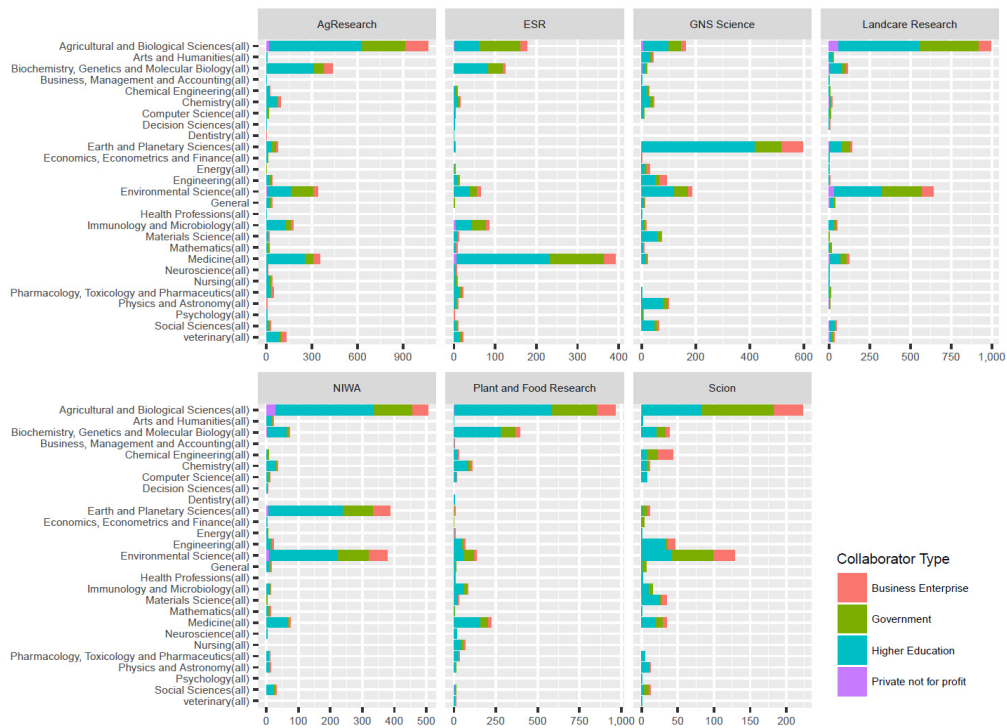

**Comparison of collaboration records of 7 Crown research institutes based on ASJC field**